\documentclass[aps,prl,twocolumn,showpacs]{revtex4-1}
\usepackage{amsbsy,latexsym,amsmath}
\usepackage{amsfonts}
\usepackage{amssymb}
\usepackage[mathscr]{eucal}
\usepackage{epsfig,graphics,graphicx}

\newcommand{\ket}[1]{\vert#1\rangle}
\newcommand{\bra}[1]{\langle#1\vert}
\newcommand{\braket}[2]{\langle#1\vert#2\rangle}

\newcommand{\expo}[1]{\mathrm{e}^{#1}}
\newcommand{\abs}[1]{\left|{#1}\right|}

\begin{document}
\title{Quantum Metrology Assisted with Abstention}
\author{B. Gendra, E. Ronco-Bonvehi, J. Calsamiglia, R. Mu\~{n}oz-Tapia, and E. Bagan}
\address{F\'{i}sica Te\`{o}rica: Informaci\'{o} i Fen\`{o}mens Qu\`antics, Departament de F\'{\i}sica, Universitat Aut\`{o}noma de Barcelona, 08193 Bellaterra (Barcelona), Spain}

\begin{abstract}

The main goal of quantum metrology is to obtain accurate values of physical parameters using quantum probes. In this context, we show that abstention, i.e., the possibility of getting an inconclusive answer at readout, can drastically improve the measurement precision. We focus on phase estimation and quantify the required amount of abstention for a given precision. We also develop analytical tools to obtain the asymptotic behavior of the precision and required rate of abstention for arbitrary pure qubit states.

\end{abstract}

\maketitle

In its simplest form, a  quantum metrology  problem has the following structure. A quantum system undergoes some physical interaction determined by some continuous parameters.
The value of the parameters gets imprinted onto the evolved state and the task of the metrologist consists  in uncovering it.
For this purpose, she performs a 
suitable measurement on the system and, based on its  outcome, she  produces a guess for the unknown value of the parameters as accurate as possible~\cite{giovannetti}. The overall performance of  the whole procedure can be
quantified by the average of a figure of merit (typically the fidelity) over some a priori distribution of the parameters and over all possible outcomes.
Phase estimation with pure states of~$N$ qubits is a paradigmatic example of this endeavor and will serve as the
main exemplification of our findings. 

In standard parameter estimation protocols~\mbox{\cite{holevo,helstrom,massar-popescu,bagan-optimal}}  the experimentalist is expected to produce a conclusive answer (maybe not right or accurate enough),  at each run of the experiment. Here we show that there are situations
where the ultimate precision of the standard approach can be  improved substantially
 if one allows for a number of inconclusive responses, where the metrologist abstains from producing a guess. This is especially relevant in situations where she can afford to re-run the experiment (i.e. she can easily prepare a new instance of the problem) or where she prioritises having high-quality estimates. Abstention has already been used in the context of state discrimination~\cite{unambiguous,fiurasek,sugimoto,bagan-q}, where some fixed rate~$Q$ of inconclusive outcomes can lower the probability of error significantly (even down to zero, as in unambiguous discrimination~\cite{unambiguous}) and can be seen as a particular example of post-selection~\cite{massar-popescu2}. 

In the estimation framework the effects of abstention have hardly  been considered.
 In the cases studied so far in the literature~\cite{fiurasek2,gendra}, abstention has limited impact for  large samples unless the experimentalist refrains from producing an estimate most of the time.
 Here we show that there are relevant cases where abstention has a dramatic effect.  Though we focus on phase estimation, other problems, such as direction estimation, can be tackled in the same way (results will be given elsewhere~\cite{in-preparation}).

In this letter, we also present a very general technique to obtain the asymptotic form of pure-state parameter estimation problems, with or without abstention. The main idea is that the components of the probe state can be viewed as a discretization of some continuous function $\varphi(t)$ on the unit interval $[0,1]$,
and the maximization of the fidelity as a constrained variational problem defined by a suitable action for $\varphi(t)$, from which the solution can be worked out analytically in many physically relevant situations.
It is worth mentioning that even in the standard approach to estimation (without abstention) analytical asymptotic expressions were known in just a few cases. 

 In phase estimation, one aims to optimally estimate the parameter $\theta$ of 
 the covariant family of states~\mbox{$\{|\Psi(\theta)\rangle=U(\theta)|\Psi_0\rangle\}_{\theta\in[0,2\pi)}$}, where~$U(\theta)$~stands for the unitary transformation~$U(\theta)|j\rangle={\rm e}^{i\theta j}|j\rangle$, and $|\Psi_0\rangle$ is a normalized fiducial state that in the eigenbasis of $U(\theta)$ can be written as $|\Psi_0\rangle=\sum_{j=0}^N c_j |j\rangle\in ({\mathbb C}^{2})^{\otimes N}$.  
The fidelity between the true phase $\theta$ and its estimate~$\theta_\chi$ can be written as~\mbox{$[1+\cos(\theta-\theta_\chi)]/2$}, where $\chi$ is a subscript specifying that the estimate is based on the outcome~$\chi$ of a generalized measurement. This is  characterized mathematically by a positive operator valued measure (POVM) $\Pi=\{\Pi_\chi\}\cup\{\Pi_0\}$,
$\Pi_0+\sum_{\chi}\Pi_\chi=\openone$, where $\Pi_\chi\geq 0$ outputs a conclusive answer (from which an estimate is proposed) and $\Pi_0\geq 0$ outputs `abstention'.
The probability of abstention taking place is
\begin{equation}\label{def Q}
Q=\int {d\theta\over 2\pi}\, \langle\Psi(\theta)|\Pi_0|\Psi(\theta)\rangle,
\end{equation}
and $\bar Q=1-Q$ is the acceptance probability (rate at which we provide definite estimates).
The average fidelity for this rate of abstention is
\begin{equation}
F(Q)\kern -0.2em ={1\over \bar Q}\sum_\chi\!\!\int \!{d\theta\over 2\pi} {1+\cos( \theta-\theta_\chi)\over2} \,\langle\Psi(\theta)|\Pi_\chi|\Psi(\theta)\rangle.
\label{eq:fidq}
\end{equation}
%

%
%
Estimation with abstention can be reduced to a standard estimation problem by simply introducing the new POVM $\tilde\Pi$, with elements given by
$\tilde\Pi_\chi\equiv\left(\openone-\Pi_0\right)^{-1/2} \Pi_\chi  \left(\openone-\Pi_0\right)^{-1/2} $,
and the new family of (normalized) states
\begin{equation}
\left\{|\tilde\Psi(\theta)\rangle\equiv{\left(\openone-\Pi_0\right)^{1/2} \over\bar Q^{1/2}}|\Psi(\theta)\rangle\right\}_{\theta\in\left[0,2\pi\right)}.
\end{equation}

%
This formulation brings forward an interpretation of the role of abstention here
:  each  initial state  $|\Psi(\theta)\rangle$ is transformed into a new $|\tilde\Psi(\theta)\rangle$ that `encodes' the unknown parameter $\theta$  in a more efficient way.  
This map improves the estimation precision by effectively increasing the distinguishability between the signal states, therefore it can only be implemented in a probabilistic fashion (it succeeds with probability~$\bar Q$). This stochastic map is fully specified  by the optimal choice of $\Pi_0$, i.e., the one that maximizes~\eqref{eq:fidq}.
%
%
Although this may seem a difficult optimization problem, a huge simplification arises because of the covariance of the family of states. Already from Eqs.~(\ref{def Q}) and \eqref{eq:fidq} one can 
easily  see  that the optimal POVM can be chosen to be covariant under the set of unitaries~$\{U(\theta)\}$. In particular this means that $\Pi_0$  can be taken invariant under the transformations $\{U(\theta)\}$.
Thus,  using Shur's lemma one gets
$\Pi_0=\sum_j f_j |j\rangle\langle j|$,  and the maximization
 is over $\{f_j\,:\, 0\le f_j\le 1\}_{j=0}^N$. Note that the transformed set of states $\{|\tilde\Psi(\theta)\rangle\}$ is also a covariant family, just as the original one, with fiducial state 
\begin{equation}\label{transformed}
|\tilde\Psi_0\rangle=\sum_{j=0}^N{ c_j\sqrt{\bar f_j}\over \sqrt{\bar Q}}|j\rangle=
\sum_{j=0}^N\xi_j|j\rangle \equiv\ket{\xi} ,
\end{equation} 
where $\bar f_j\equiv 1-f_j$.
%

Since the transformed states are still covariant, we can choose $\tilde \Pi$ to be the well known optimal continuous and covariant POVM \cite{holevo,bagan-optimal}: $\{\tilde\Pi(\theta)=U(\theta)|\Phi\rangle\langle\Phi| U^\dagger(\theta)\}_{\theta\in\left[0,2\pi\right)}$, where  $|\Phi\rangle=\sum_{j=0}^N|j\rangle$.
Hereafter it is assumed that the states have non-negative coefficients  $\xi_{j}\geq 0$ (and hence $c_{j}\geq 0$), for any phases present in the coefficients $c_{j}$ (or~$\xi_{j}$) can be absorbed by the above POVM elements. With this, the calculation of the fidelity 
simplifies to
%
\begin{equation}\label{fidelity-general-3}
F=\frac{1}{2}+\frac{1}{2} \bra{\xi} \mathsf{M}  \ket{\xi},
\end{equation}
where in the eigenbasis of $U(\theta)$ 
the matrix of~$\mathsf{M}$ is real and tridiagonal, with elements $M_{ij}=\left( \delta_{i,j+1}+\delta_{i,j-1}\right)/2$.
%
 %
%
The maximization over~$\{f_j\}$ in~(\ref{fidelity-general-3}) can be turned into a maximization over the transformed states,~\mbox{$\Delta\equiv\max_{\ket{\xi}} \bra{\xi} \mathsf{M}  \ket{\xi}=\max_{\{\xi_j\}} \sum_{j=0}^{N-1} \xi_j\xi_{j+1}$}, 
subject to the constraints
$\braket{\xi}{\xi}=1$ and $\xi_j \leq\lambda c_j$, where $\lambda\equiv\bar{Q}^{-1/2}$.
Then, the maximum fidelity for a given rate of abstention~$Q$ is~$F(Q)=(1+\Delta)/2$. 

For large enough abstention rates the inequality constraint has no effect (provided $c_j\neq 0$, $\forall j$) and  $\Delta$ becomes the maximum eigenvalue of the matrix $\mathsf M$. In this case, $F^{*}\equiv F(Q\to1)$ is the maximum fidelity that can be achieved by optimizing the components of the fiducial state; these are given by the corresponding eigenvector~$\ket{\xi^{*}}$ of~$\mathsf M$
\cite{fiurasek2}. 
From the inequality constraint we obtain the critical acceptance rate 
$\bar{Q}^{*}=\min_{j} (c_{j}/{\xi_{j}^{*}})^{2}$.
That is, for abstention rates such that~\mbox{$Q\ge Q^{*}= 1- \bar{Q}^{*}$} the fidelity attains its absolute maximum value $F^{*}$. 
In the other extreme, when no abstention is allowed \mbox{($Q=0$)}, no maximization is possible, hence $\xi_j= c_j$ and~\mbox{$\Delta=\langle c|{\mathsf M}|c\rangle$}.
For intermediate values of $Q\in(0,Q^{*})$ the calculation becomes more tricky. 
For moderate values of $N$ 
it can be easily cast as a semidefinite programming (SDP) problem (see Supplemental Material), and hence solved efficiently to arbitrary accuracy~\cite{KKT}.

However, the main focus of this work is on asymptotics (large $N$ regime) and in particular on presenting an approach that enables obtaining analytical asymptotic expressions.
%
%
With this aim, let us define $S\equiv 1-\bra{\xi}{\mathsf M}\ket{\xi}$, where we further introduce Lagrange multipliers $b^2/2$ for the normalization condition and $s_j$, $j=0,\dots,N$ for the 
inequality constraints $\xi_j \leq\lambda c_j$, also called  {\em primal feasibility} conditions. We thus have to minimize
\begin{eqnarray}
S&=&
\frac{1}{2}\left[\sum_{j=0}^{N-1}\left(\xi_{j+1}-\xi_j\right)^2+\xi_1^2+\xi_{N}^2\right]\nonumber\\
&-&{b^2\over2}\left(\sum_{j=0}^N\xi_j^2-1\right)+\sum_{j=0}^N s_j\left(\xi_j-\lambda c_j\right) ,
\label{eq:action}
\end{eqnarray}
where the so-called
{\em dual feasibility} conditions
~\mbox{$s_j\ge0$},
and the   {\em complementary slackness} conditions~\mbox{$ s_j\left(\xi_j-\lambda c_j\right)=0 $} must also be imposed, as dictated by the Karush-Kuhn-Tucker  (KKT) method (see e.g.,~\cite{KKT}).
%


Rather than attempting to solve the above minimization, 
we will reframe it as a functional variational problem by taking $N$ to be asymptotically large. 
%
We first note that as $N$ goes to infinity $j/N$ approaches a continuous real variable~$t$. So, we define
$0\le t\equiv{j/N}\le 1$,
and assume $\{\xi_j\}$ and $\{c_j\}$ are a discretization of some continuous functions, $\varphi(t)=\xi_j\sqrt{N}$
and $\psi(t)=c_j\sqrt{N}$.
%
%
Note  that $\varphi(t)\ge0$ and~$\psi(t)\ge0$, and the normalization condition $\int_0^1 \varphi(t)^2=\int_0^1 \psi(t)^2=1$ is satisfied.
%
%
%
%
%
%
{}It follows that $\xi_{j+1}-\xi_j\simeq N^{-3/2} [d\varphi(t)/dt]$, and we can write \eqref{eq:action} as the functional
\begin{eqnarray}
&&\kern-1emS[\varphi]=\!{\varphi^2(0)+\varphi^2(1)\over 2N}\nonumber \\[.5em]
&&\kern-1em\phantom{S}+\!{1\over N^2}\!\!\int_0^1 \!\!dt\!\left[{1\over2 }\!\left(d\varphi\over dt\right)^2\!\!\!-{\omega^2\over2}\!\left(\varphi^2\!-\!1\right)\!+\!\sigma(\varphi-\lambda\psi)\right]\!,
\label{ebc13.06.12-1}
\end{eqnarray}
where $\omega=N b\geq 0$ is the properly scaled Lagrange multiplier and $\sigma(t)$ is a function that interpolates the set of multipliers $\{s_j\}$, i.e., $s_j= N^{-5/2}\sigma(t)$.
With this, the following conditions must hold: \mbox{$\varphi(t)-\lambda\psi(t)\phantom{]}\le0$} ({\rm primal feasibility}),  ~\mbox{$\sigma(t)\ge 0$} ({\rm dual feasibility}),
and ~\mbox{$\sigma(t)[\varphi(t)-\lambda\psi(t)]=0$} ({\rm complementary slackness}).
Note that by imposing the boundary conditions $\varphi(0)=0$ and $\varphi(1)=0$,
the functional $S[\varphi]$ becomes $O(N^{-2})$. These conditions can be always met if $Q>0$,  while for $Q=0$ the asymptotic scaling of~$F$ is dominated  by the first 
term of ~\eqref{ebc13.06.12-1}, of order $N^{-1}$.

More interestingly, the minimization of $S[\varphi]$ defines a mechanical problem, of which the second line in~Eq.~(\ref{ebc13.06.12-1}) is the  `action' and the corresponding integrant the `Lagrangian'.
%
%
It describes a driven harmonic oscillator with angular frequency $\omega$,
whose `equation of motion' is
\begin{equation}
{d^2\varphi\over dt^2}+\omega^2\varphi=\sigma .
\label{dho}
\end{equation}
To solve this problem, we first note that the
slackness conditions imply that either $\varphi(t)=\lambda\psi(t)$, in which case we say~$t$ is in the so called {\em coincidence set} $\mathscr C$, or  $\sigma(t)=0$. In the second  case, $t\in\bar{\mathscr C}$ (the complement of~$\mathscr C$), the primal feasibility condition is~$\varphi(t)<\lambda\psi(t)$ and Eq.~(\ref{dho}) becomes homogeneous. It has the familiar solution
~\mbox{$
\varphi(t)=A\sin\omega t+ B \cos\omega t ,\label{sho-no}
$}
where $A$, $B$ and $\omega$ are constants to be determined.
In 
$\mathscr C$,~$\sigma$ is given by~\eqref{dho}, where we make the substitution~$\varphi(t)=\lambda\psi(t)$. 
If we restrict ourselves to fiducial states $|\Psi_0\rangle$ whose  components $c_j$ are such that $\psi(t)$ 
is continuous in the whole unit interval, one can show that the solution $\varphi(t)$ and its first derivative must be also continuous there [except in points of $\mathscr C$ where $\psi(t)$ itself is not differentiable]. Most of the physically relevant cases are of this type. By taking into account the boundary conditions, as well as the `matching conditions', namely, continuity of $\varphi(t)$ and its derivative in the boundaries of $\mathscr C$, one can determine the arbitrary constants that arise in solving the equation of motion, which include the location of the boundaries of~$\mathscr C$. 
%
%
%
%
The minimum value of $S$  for $Q>0$ can be expressed in terms of the Lagrange multiplier (function) $\omega$ ($\sigma$), and the given function $\psi$, as
%
%
\begin{equation}
S={1\over N^2}\left({\omega^2\over2}-{\lambda\over2}\int_0^1 dt\,
\sigma\psi\right)  .
\label{action}
\end{equation}
%
Note that the integral is effectively over 
$\mathscr C$, where $\sigma=\lambda (d^2\psi/dt^2+\omega^2 \psi)$. 
%
In the following we solve some relevant cases, but we first compute the fidelity for arbitrary large rates of abstention, asuming $c_j\neq 0$, $\forall j$. 

\emph{Large abstention:} 
For  abstention rates very close to one (large~$\lambda$), one has 
${\mathscr C}=\emptyset$, and the above reduces to a maximum eigenvalue problem. 
The solution is
~\mbox{$
\varphi(t)=\sqrt2\sin\pi t.
$}
%
%
%
%
This  yields the asymptotic result 
\begin{equation}
F^{*}=1-{\pi^2\over4N^2} ,
\label{ebc29.04.12-3}
\end{equation}
which coincides with the known maximum fidelity for optimal phase encoding~\cite{BMM-T,fiurasek2}.

\emph{Phase states}: The elements of this family are generated by an equal superposition of all Fock states $\ket{j}$, $ c_j=1/\sqrt{N+1}$ (hence proportional to the POVM seed state). The corresponding continuous version is
$\psi(t)=1$. 
Without abstention,~$Q=0$ ($\lambda=1$), the optimal phase estimation precision provided by these states does not exceed the shot-noise limit: $1-F=1/(2N+2)$. 
%
%
For  $Q>0$ ($\lambda>1$) the situation changes markedly. We can freely impose~$\varphi(0)=\varphi(1)=0$ and get rid of the shot-noise type of term $1/N$ in  \eqref{ebc13.06.12-1}.
In a sufficiently small neighbourhood  of $t=0$, 
we have $\varphi(t)-\lambda<0$, and hence $\sigma(t)=0$ there. If $\alpha$ is the maximum value of $t$ less than $1/2$ for which this condition holds, it must be a boundary point of $\mathscr C$.  Then, for
 $\alpha\leq t\leq 1/2$ the solution is given by the rescaled input state $\varphi(t)=\lambda\psi(t)=\lambda$. Thus,
%
$
\varphi(t)=
A\sin \omega t$ for $t \in[0,\alpha)$  and $\varphi(t)= \lambda$ for  $t \in[\alpha,1/2]$. 
Imposing the  matching conditions we find $\alpha=1-1/\lambda^2=Q$, $\lambda\leq\sqrt{2}$ (i.e. $Q^*=1/2$), and
%
%
%
\begin{equation}
\varphi(t)\!=\!\left\{\!\!
\begin{array}{lrcl}
\displaystyle {\bar Q^{-{1\over 2}}}\sin{\pi t\over2Q} \; &0\!&\le t<&\! \displaystyle {Q},
\\[1em]
{\bar Q^{-{1\over 2}}} &\displaystyle  {Q}&\!\le t\le\!& \displaystyle {1/2}.
\end{array}
\right.
\end{equation}
For $1/2<t\le 1$ the solution follows from the obvious symmetry relation $\varphi(t)=\varphi(1-t)$.
%
We have $\mathscr C=[Q,\bar Q]$ and $\sigma(t)=\omega^2\lambda$ for $t\in{\mathscr C}$.
Therefore, Eq.~(\ref{action}) leads to
%
%
\begin{equation}
F=1-{\pi^2\over16\,Q\bar QN^2} ,\qquad 0<Q\le Q^{*}=1/2, \label{deltaCconst2}
\end{equation}
and for $1/2<Q \le1$ we have $F=F^*$ in Eq.(\ref{ebc29.04.12-3}).
We note that even the slightest abstention rate unlocks the probing capabilities of the phase states and drastically changes the estimation precision from shot-noise ($1/N$) to Heisenberg ($1/N^{2}$) scaling.  

%
\emph{Multiple copies:} These probe states have been widely considered in quantum metrology.
They have the form
\begin{equation}\label{paral}
\ket{\Psi(\theta)}=\left(\frac{\ket{0}+\expo{i \theta}\ket{1}}{\sqrt{2}}\right)^{\otimes N} ,
\end{equation} 
%
and the coefficients of the corresponding fiducial state $\ket{\Psi_0}$ 
 read $c_j=2^{-N/2}\binom{N}{j}\raisebox{.7em}{}^{-1/2}$.
%
%
Their maximum precision  for $Q=0$ and large $N$ is known to be
$1-F=1/(4N)$~\cite{BMM-T}. 

In the asymptotic limit the rescaled components $\sqrt{N}c_j$ approach the function
\begin{equation}
\psi(t)=\left[{N\over 2\pi t(1-t)}\right]^{1/4}\exp\left\{-\frac{N}{2}H(t\parallel 1/2)\right\},\label{ccks}
\end{equation}
where $H(t\!\parallel\! 1/2)=\log 2+t\log t+(1-t)\log(1-t)$ is the relative entropy between two Bernoulli distributions with success probabilities $t$ and $1/2$.
The symmetry of the problem suggests using the variable~\mbox{$\tau=t-1/2$}, $\tau\in[-1/2,1/2]$. As a function of this new variable $\varphi(\tau)$ must be even [we slightly abuse notation by writing $\varphi(t(\tau))$ in short as $\varphi(\tau)$], hence it must have the form
 %
\begin{equation}
\varphi(\tau)=\left\{
\begin{array}{ll}
A\cos \omega\tau ,&\quad 0\le \abs{\tau}\le\alpha,\\[.5em]
\lambda\psi(\tau),&\quad \alpha<\abs{\tau}\le 1/2,
\end{array}
\right.
\label{sol2}
\end{equation}
where as above $\omega$, $\alpha$ and $A$ are determined by the matching and normalization conditions. 
The resulting equations cannot be solved in full generality, however we will obtain analytical solutions in two relevant  regimes.  
To this end, we note that in the region~$|\tau|\lesssim N^{-1/2}$ [i.e., around the peak at $t=1/2$ of~(\ref{ccks})], $\psi(\tau)$ behaves as the Gaussian distribution $\psi(\tau)\!\approx\!\left(2N /\pi\right)^{1/4}\!\expo{-N\tau^2}\!\!$,
%
 %
whereas at the tails ($|\tau|>N^{-1/2}$) it falls off at an exponential rate lower bounded by $H(1/2+\alpha\!\parallel \!1/2)/2$.

In the first regime, the abstention rate $Q$ has a fixed non-zero value (independent of $N$).  
This requires that the boundary point $\alpha$ scales
as the width of   $\psi(\tau)$,  i.e., $\alpha\sim N^{-1/2}$,
thus $\psi(\tau)$ is accurately given by its Gaussian 
approximation. This observation, along with a suitable 
rescaling of $\omega$ and $A$, enable us to
obtain the solution in parametric form: $(Q(\Omega), S(\Omega))$ (see Supplemental Material). In Fig.~\ref{SvsQ2}~(a) we plot $N S=2N(1-F)$ as a function of $Q$. Despite the strong dependence on the abstention rate, particularly at $Q\lesssim1$, the precision will be  shot-noise limited in the whole interval $[0,1)$. We see that, e.g., an abstention of about $90\%$ has the same effect as doubling the number of copies in the standard approach, with $Q=0$. Also, for vanishing abstention rate we recover the well known result~\mbox{$2N(1-F)=1/2$}. The exact (numerical) profile of the transformed  state $|\tilde\Psi_0\rangle$ is shown in Fig.~\ref{SvsQ2}~(b), together with the analytical result $\varphi(t)$, for an abstention rate of $56\%$ and two different values of $N$ of $20$ and $80$ copies.
\begin{center}
\begin{figure}
	\centering
	\setlength{\unitlength}{5mm}
\thinlines
\begin{picture}(15.3,7.3)(0,0)
\put (-.6,1.3){\includegraphics[scale=.47]{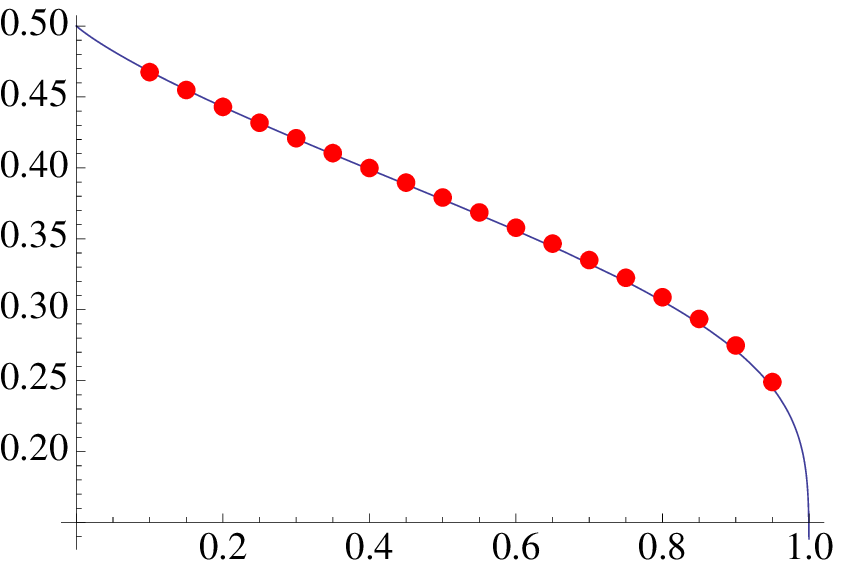}}
\put (7.778,1.28){\includegraphics[scale=.5]{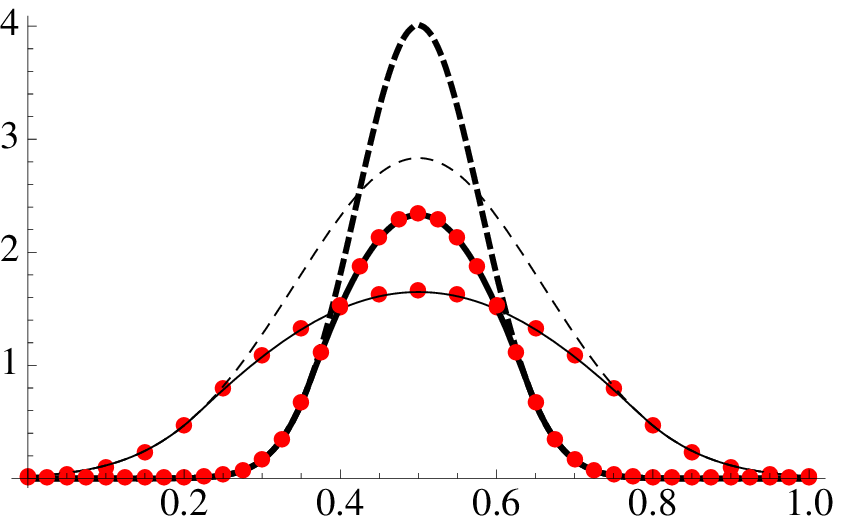}}
\put (6.8,3.7){\rotatebox{90}{\scriptsize$\varphi(t)$, $\sqrt{N}\xi_j$}}
\put (14.5,.6){\scriptsize$t$, $j/N$}
\put (-1.2,5.7){\rotatebox{90}{\scriptsize$N S$}}
\put (6.5,.6){\scriptsize$Q$}
\put (5.5,6){(a)}
\put (15.,6){(b)}
\end{picture}
\caption{(Color online) (a) Plot of $NS=2N(1-F)$ vs $Q$ (solid line) for an asymptotically large number $N$ of parallel spins, Eq.~(\ref{paral}). The dots have been obtained by numerical optimization with $N=100$. (b) Profile of the transformed fiducial state $|\tilde\Psi_0\rangle$ 
for~\mbox{$\lambda=1.5$}  ($Q=0.56$). The thin (thick) line corresponds to $N=20$ ($N=80$). The points are obtained by numerical optimization. The dashed lines represent the function~$\lambda\psi(t)$ that defines the primal feasibility condition $\varphi(t)\le\lambda\psi(t)$, where $\psi(t)$ is in~(\ref{ccks}). }\label{SvsQ2} 
\end{figure}
\end{center} 
\vspace{-2.1em}

We next explore a second regime in which the Heisenberg limit $1-F\sim N^{-2}$ (i.e., $S\sim N^{-2}$) is attained.
For that we need that the  function  $\varphi(\tau)$ in \eqref{sol2} has a wide (non vanishing) 
support as $N$ goes to infinity. This is easily accomplished by taking the   boundary point $\alpha$ to be independent of $N$. 
In this situation, once $N$ is large enough, the coincidence set~$\mathscr C$ lies on the tails of~$\psi(\tau)$, where this function falls off exponentially at the dominant rate of~$H(1/2+\alpha\!\parallel \!1/2)/2$. Solving the matching and normalization conditions we obtain the fidelity: 
 \begin{equation}
 F=1-\frac{\pi^2}{16N^2\alpha^2}+O(N^{-3});\quad 0<\alpha\le1/2;
 \label{fid-heis}
 \end{equation}
 and the  acceptance rate: {$Q\sim \exp\{-N H(1/2+\alpha\!\parallel \!1/2)\}$}.
 Thus, the Heisenberg limit scaling $1/N^2$ can be attained if such exponential rate of acceptance is affordable. Note that the critical acceptance rate is $\bar Q^*=2^{-N}$ ($\alpha\to1/2$), below which $F=F^*$. 

%

Although in this letter we have focused in phase estimation, our methods directly apply to other problems such as direction estimation~\cite{in-preparation}. In this case one can define the analogous of the phase states introduced above (fiducial state proportional to the seed of the corresponding optimal POVM).  Our results nicely mirror those presented here, the main difference being the replacement of the trigonometric functions by Bessel functions. 
For tensor product states of identical copies,  abstention does not improve the estimation \cite{gendra}. However, for antiparallel product states where half of the spins point in one direction while the other half  point in the opposite,
abstention again gives raise to a  change in the  asymptotic behavior of the precision, from the original shot-noise ($1-F\sim 1/N$) to the Heisenberg  ($1-F\sim 1/N^2$) scaling.

We thank G. Chiribella for his contributions at the early stages of this work. We acknowledge financial support from ERDF: European Regional Development Fund. This research was supported by
 the Spanish MICINN, through contract FIS2008-01236 and the Generalitat de
Catalunya CIRIT, contract  2009SGR-0985.

\section{Supplemental Material}
\subsection{Semidefinite programming formulation}
The evaluation of
\begin{equation}
\Delta\equiv\max_{\ket{\xi}} \bra{\xi} \mathsf{M}  \ket{\xi}
\label{frivolous1}
\end{equation} 
subject to $\braket{\xi}{\xi}=1$ and $\xi_j \leq\lambda c_j$, can be  cast as a semidefinite programming (SDP) problem. 
One simply linearizes these constraints  by introducing a SDP (positive operator) variable $\mathsf B$ to play the role of $\ket{\xi}\!\bra{\xi}$. The SDP form of~(\ref{frivolous1}) is then
\begin{equation}\label{Delta AMA sdp}
\Delta\equiv\max_{\mathsf B}  \,{\rm tr}\left(\mathsf{M}  \mathsf{B}\right)
\end{equation}
subject to
\begin{eqnarray}\label{sdp 2}
&\;\;\;&{\rm tr}\, \mathsf{B}=1, \;\;\; \mathsf{B}\geq 0, \nonumber\\
&& \mathsf{B}_{jj}\leq\lambda^2 |c_j|^2,\quad \lambda\ge 1.
\end{eqnarray}
In order for the optimization problems~(\ref{frivolous1})  and~(\ref{Delta AMA sdp}) to be equivalent, we must prove that the optimal $\mathsf B$ is rank one. {\em Proof}\,:  since all the entries of~$\mathsf M$ are non-negative, ${\rm tr}({\mathsf M}{\mathsf B})$ increases with increasing values of the off-diagonal entries~$\mathsf B_{j,j+1}$.  Their maximum value consistent with positivity is given by rank one matrices.
Therefore, the optimal $\mathsf B$ is of the form $\ket{\xi}\!\bra{\xi}$. 


\subsection{Shot-noise regime for multiple copies}


As explained in the main text, this regime requires that $\alpha\sim N^{-1/2}$ to ensure that the scaled components~$\sqrt{N}c_j$ of the fiducial estate $|\Psi_0\rangle=|\Psi(0)\rangle$, Eq.~(\ref{paral}),
can be approximated by the Gaussian distribution
\begin{equation}
\psi(\tau)\approx\left(\frac{2N }{\pi}\right)^{\!\!1/4}\!\!\expo{-N\tau^2} \label{gaussc}
\end{equation}
at the boundary of $\mathscr C$. Asymptotically, the scaled components $\sqrt N\xi_j$ of the optimal $|\xi\rangle$ in~(\ref{frivolous1}) approach the function
%
\begin{equation}
\varphi(\tau)=\left\{
\begin{array}{ll}
A\cos \omega\tau ,&\quad 0\le \abs{\tau}\le\alpha,\\[.5em]
\lambda\psi(\tau),&\quad \alpha<\abs{\tau}\le 1/2,
\end{array}
\right.
\label{sol2p}
\end{equation}
[Eq. (15) in the main text)]. %
The matching conditions, i.e., the continuity of both $\varphi(\tau)$ and $\varphi'(\tau)$ at the boundary point $\tau=\alpha$ can be combined to obtain
%
%
%
%
%
%
\begin{eqnarray}
a^2&=&\frac{\Omega\tan\Omega}{2},\label{as} \\[.5em]
A^2&=&\left({2N\over\pi}\right)^{1/2}\frac{\lambda^2\expo{-2a^2}(4a^4+\Omega^2)}{\Omega^2} , \label{A0} 
\end{eqnarray}
where we have defined $\Omega\equiv\omega\alpha$. The normalization condition turns out to be

\begin{eqnarray}
1&=&A^2\frac{a (2 \Omega+\sin2\Omega)}{2\sqrt N\,\Omega}\!+\!\lambda^2\left[1\!-\!{\rm Erf}(\sqrt{2}\,a)\right]\!,\label{norm3}
\end{eqnarray}
where ${\rm Erf}(x)$ is the error function.
Eq.~(\ref{norm3}) is correct up to exponentially vanishing contributions, which can be neglected here.
In deriving this equation  we also used that~${\rm Erf}(\sqrt{N/2})\to 1$ for large $N$. 
%
%
Substituting Eq.~(\ref{A0})  in Eq.~(\ref{norm3}) we obtain
\begin{equation}
{1\over \lambda^2}\!=\!{\rm Erfc}(\sqrt{2} a)\!+\!\frac{
 a  (4 a^4 \!+\! \Omega^2) (2 \Omega \!+ \!
    \sin2\Omega)}{\sqrt{2 \pi} \Omega^3}\expo{-2 a^2}\!,
    \label{1/lambda2}
 \end{equation} 
 where $\rm Erfc$ is the complementary error function, defined as ${\rm Erfc}(x)=1-{\rm Erf}(x)$.
 %
%
%
Finally, with the help of the Gaussian approximation~(\ref{gaussc}), we compute the minimum action, Eq.~(\ref{action}), and obtain
\begin{equation}\label{S equator}
 S\!=\!{\omega^2\over2N^2}-{\lambda^2\over2N^2}\!\!\left[
(\omega^2\!\!-\!N){\rm Erfc}(\sqrt{2} a)\!+\!{4N^{\phantom{3\over2}}\kern-.6em a\over\sqrt{2\pi}}{\rm e}^{-2 a^2}
\!\right]\!\!.
\end{equation}
Eqs.~(\ref{as}) and~(\ref{1/lambda2}), along with $\omega=\Omega\,\sqrt N/a$ and $Q=1-1/\lambda^2$, enable us to write all variables in terms of the single parameter $\Omega$. By further substituting in Eq.~(\ref{S equator}) we obtain the curve $(Q,S)$ in parametric form:
 \begin{widetext}
 \begin{eqnarray}
  Q&=&{\rm Erf}\left(\sqrt{\Omega\tan\Omega}\,\right)-\left(\Omega\sec^2\Omega+\tan\Omega\right)\sqrt{{\tan\Omega\over\pi\,\Omega}}\,\expo{-\Omega\tan\Omega},\label{para1}\\  
  S&=&{1\over2 N}\left[1 + \frac{\tan^2\Omega-\Omega\left(2\Omega-\tan\Omega\right)\sec^2\Omega}{2\,\Omega^2\sec^2\Omega+\sqrt{\pi\,\Omega\tan\Omega}\;{\rm Erfc}\left(\sqrt{\Omega\tan\Omega}\right)\,\expo{\Omega\tan\Omega}}\right]^{-1} .\label{para2}
 \end{eqnarray}
 \end{widetext}
 This curve, which can be easily plot, is shown in Fig.~\ref{SvsQ2}~(a).
An analytic expression of $S$ as an explicit function of~$Q$ cannot be found  since that would  involve solving the non-trivial transcendental equation in~(\ref{para1}).


\begin{thebibliography}{99}
\newcommand{\jPRL}[3]{Phys.~Rev. Lett.~\textbf{#1}, #2~(#3)}
\newcommand{\jPRA}[3]{Phys.~Rev. A~\textbf{#1}, #2~(#3)}
\newcommand{\jPRD}[3]{Phys.~Rev. D~\textbf{#1}, #2~(#3)}
\newcommand{\jJPA}[3]{J.~Phys. A~\textbf{#1}, #2~(#3)}
\newcommand{\jPLA}[3]{Phys.~Lett. A~\textbf{#1}, #2~(#3)}
\newcommand{\jNJP}[3]{New.~J.~Phys.~\textbf{#1}, #2~(#3)}
\newcommand{\jJOB}[3]{J.~Opt. B~\textbf{#1}, #2~(#3)}
\newcommand{\jJMP}[3]{J.~Math.~Phys.~\textbf{#1}, #2~(#3)}
\newcommand{\jJMO}[3]{J.~Mod.~Opt.~\textbf{#1}, #2~(#3)}
\newcommand{\jRMP}[3]{Rep.~Math.~Phys.~\textbf{#1}, #2~(#3)}
\newcommand{\jIEEE}[3]{IEEE Trans.~Inf.~Theory~\textbf{#1}, #2~(#3)}
\newcommand{\jNPHOT}[3]{Nature Photonics~\textbf{#1}, #2~(#3)}
\bibitem{giovannetti} V.~Giovannetti, S.~Lloyd and L.~Maccone, \jNPHOT{5}{222}{2011}.
\bibitem{holevo} A.S.~Holevo, {\it Probabilistic And Statistical Aspects Of Quantum Theory},
 North-Holland Series In Statistics And Probability (North-Holland, Amsterdam 1982).
 

 %
 \bibitem{helstrom}C. W. Helstrom, \emph{Quantum Detection and Estimation Theory} (Academic Press, New York, 1976).
\bibitem{massar-popescu} S.~Massar and S.~Popescu, \jPRL{74}{1259}{1995}.

\bibitem{bagan-optimal} E.~Bagan, M.~Baig, A.~Brey, R.~Munoz-Tapia, and R.~Tarrach,
\jPRL{85}{5230}{2000}; \jPRA{63}{052309}{2001}.                                                                           
  

\bibitem{unambiguous}I. D. Ivanovic, \jPLA{123}{257}{1987}.

 \bibitem{fiurasek}J. Fiur\'a\v{s}ek and M. Je\v{z}ek, \jPRA{67}{012321}{2003}.

\bibitem{sugimoto}H. Sugimoto, T. Hashimoto, M. Horibe, and A. Hayashi, \jPRA{80}{052322}{2009}.

 
 \bibitem{bagan-q} E.~ Bagan, R.~Munoz-Tapia, G.A.~Olivares-Renteria and J.A.~Bergou, 	arXiv:1206.4145v1 [quant-ph].                 

 \bibitem{massar-popescu2} S.~Massar and  S.~Popescu, \jPRA{84}{052106}{2011}.


\bibitem{fiurasek2} J.~Fiur\'a\v{s}ek, \jNJP{8}{192}{2006}.


%

\bibitem{gendra} B.~Gendra, E.~Ronco-Bonvehi, J.~Calsamiglia, R-~Munoz-Tapia, and E.~Bagan, arXiv:1205.5479v1 [quant-ph]. To appear in New J. of Phys.

\bibitem{in-preparation} B.~Gendra, E.~Ronco-Bonvehi, J.~Calsamiglia, R-~Munoz-Tapia, and E.~Bagan, in preparation.



\bibitem{KKT} S.~Boyd and L.~Vandenberghe, {\it Convex Optimization}  (Cambridge University Press, 2004).

\bibitem{BMM-T}E.~Bagan, A.~Monras and R.~Munoz-Tapia, \jPRA{71}{062318}{2005}.


\end{thebibliography}
\end{document}